\documentstyle[12pt,equation]{article}
\begin{document}
\def\to{\rightarrow}
\def\ub{\underbar}
\def\vb{\mbox{$\bar{v}$}}
\def\wt{\widetilde}
\def\ta{\mbox{$\widetilde{A}$}}
\def\ffbar{\mbox{$f \bar{f}$}}
\def\lmax{\mbox{$\lambda_{\rm max}$}}
\def\tanb{\mbox{$\tan \! \beta$}}
\def\sinb{\mbox{$\sin \! \beta$}}
\def\cosb{\mbox{$\cos \! \beta$}}
\def\ctw{\mbox{$\cos \! \theta_W$}}
\def\stw{\mbox{$\sin \! \theta_W$}}
\def\hth{\mbox{$H_3^0$}}
\def\nt{\mbox{$N_3$}}
\def\to{\rightarrow}
\def\hbt{\mbox{$\bar{H^0_3}$}}
\def\ml{\mbox{$m_{L_1^0}$}}
\def\lone{\mbox{$L_1^0$}}
\def\ltwo{\mbox{$L_2^0$}}
\def\sigav{\mbox{$\langle \sigma_{\rm eff} v \rangle$}}
\def\omeh{\mbox{$\Omega h^2$}}
\def\mlop{\mbox{$m_{L_1^\pm}$}}
\def\mltp{\mbox{$m_{L_2^\pm}$}}
\def\la{\mbox{$\lambda$}}
\def\rs{\mbox{$\sqrt{s}$}}
\def\qqbar{\mbox{$q \bar{q}$}}
\def\ben{\begin{subequations}}
\def\be{\begin{equation}}
\def\een{\end{subequations}}
\def\ee{\end{equation}}
\def\beq{\begin{eqalignno}}
\def\eeq{\end{eqalignno}}
\def\bea{\begin{eqnarray}}
\def\eea{\end{eqnarray}}
\def\epem{\mbox{$e^+ e^- $}}
\def\aem{\mbox{$\alpha_{\rm em}$}}
\renewcommand{\thefootnote}{\fnsymbol{footnote}}
\setcounter{page}{0}
\setcounter{footnote}{0}
\begin{flushright}
MADPH--95--905\\
UT--710 \\
August 1995\\
\end{flushright}
\vspace{1cm}
\begin{center}
{\Large \bf A Decisive Test of Superstring--Inspired $E(6)$ Models}\\
\vspace{1cm}
Manuel Drees\footnote{Heisenberg Fellow}\\
{\it University of Wisconsin, Dept. of Physics, 1150 University Ave., Madison,
WI53706, USA} \\
\vspace*{0.5cm}
Atsushi Yamada\\
{\it Dept. of Physics, University of Tokyo, Bunkyo--ku, Tokyo 113, Japan}\\
\vspace{1cm}
\end{center}
\begin{abstract}
We point out that in a large class of superstring--inspired $E(6)$ models,
either an \epem\ collider operating at a center--of--mass energy $\rs= 1.5$
TeV or higher must detect the pair production of charged or neutral exotic
leptons, or an \epem\ collider with $\rs \geq 300$ GeV must discover at least
one light neutral Higgs boson with invisible branching ratio exceeding 50\%.
If neither of these two signals is seen, the lightest neutral exotic lepton
would overclose the universe, and the model could be completely excluded,
independent of the values of the numerous free parameters. Future Higgs
searches might lower the energy of the \epem\ collider needed to test these
models decisively. The only assumptions we have to make are that $R-$parity is
exact, so that the lightest exotic lepton is stable if it is lighter than the
lightest neutralino, and that no $SO(10)$ singlet scalar gets a vacuum
expectation of order $10^{10}$ GeV or higher. If the second condition is
violated, the model effectively reduces to an $SO(10)$ model as far as
collider experiments are concerned.

\end{abstract}
\newpage
\pagestyle{plain}

\section*{1) Introduction}
Superstring--inspired $E(6)$ models \cite{1} contain a large number of
particles in addition to those present in the Standard Model (SM):
Superpartners of the known matter fermions and gauge bosons; scalar di-- or
lepto--quarks; extended gauge and Higgs sectors; and new ``exotic" quarks and
leptons. Indeed, this class of models can almost be considered to be a maximal
(weakly interacting) extension of the SM. It is this aspect, rather than the
by now quite tenuous connection to superstring theory \cite{2}, that keeps
interest in these models alive \cite{3}.

Unfortunately most of the new particles predicted by $E(6)$ models could be
very heavy. In the absence of a comprehensive theory of supersymmetry breaking
we are not able to give firm upper bounds on sparticle masses \cite{4}. The
masses of the new gauge bosons contained in such models can be made very large
by postulating large vacuum expectation values (vevs) for Higgs fields that
are singlets under the SM gauge group. Finally, most of the exotic fermions
reside in vector--like representations of the SM gauge group; they can
therefore also be made very heavy. In the simplest of these models, where one
requires gauge symmetry breaking to be triggered by radiative corrections
\cite{5,6}, these possibly large scales are in fact all related: The vevs of
most Higgs singlets cannot exceed the SUSY breaking scale significantly, and
the same vevs also give rise to the masses of the exotic leptons, with Yukawa
couplings of order 1 or less if the theory is to remain weakly interacting all
the way up to the scale of grand unification (GUT). This allows to derive
non--trivial relations \cite{6} between some of these masses, but
unfortunately does not exclude the possibility that they are all very large.

There are exceptions to this rule, however. Each fermion generation of $E(6)$
contains 27 degrees of freedom. Two of those are SM singlets, commonly called
$\nu_R$ and $N$. The $\nu_R$ resides in the {\bf 16} of $SO(10)$, together
with the 15 degrees of freedom that form a complete fermion generation in the
SM. The $\nu_R$ fields might be exactly massless; alternatively, they might
get very large masses through non--renormalizable operators if some scalar
$\tilde{\nu}_R$ field gets a vev at an ``intermediate" scale around 10$^{10}$
GeV or more, which could also give rise to see--saw type neutrino mass
matrices \cite{7}. Either way it is very difficult to derive significant
constraints on these $\nu_R$ fields.\footnote{Successful nucleosynthesis in
the early universe requires a large freeze--out temperature, i.e. large masses
for gauge bosons coupling to $\nu_R$, if $\nu_R$ is exactly massless \cite{8};
however, direct experimental searches \cite{3} by now constrain these gauge
bosons to be heavier than several hundred GeV anyway.}

In contrast, the $N$ superfields are singlets under $SO(10)$. Their fermionic
components can acquire masses at the weak scale only by mixing with the
neutral components of exotic $SU(2)$ doublet fermions \cite{9}; there are no
terms of the type $N^3$ in the superpotential, since $N$ is not a singlet
under the complete gauge group. Note that the vevs that give rise to this
mixing also break the $SU(2) \times U(1)_Y$ symmetry of the SM, which means
that they contribute to the masses of the $W$ and $Z$ bosons. This allows to
derive a firm upper bound \cite{10} of just over 100 GeV for the mass of the
lightest eigenstate resulting from this mixing, if the relevant Yukawa
couplings are required not to have a Landau pole below the GUT scale. This
bound can only be avoided if some $N$ scalar has a vev of order $10^{10}$ GeV
or more, and if the superfields containing the light exotics have
non--renormalizable couplings to this vev. However, the same vev would also
allow to give very large masses to all new gauge bosons and charged new matter
fermions; at scales below $\langle N \rangle$ one would then just end up with
the minimal supersymmetric standard model (MSSM) \cite{4}. We will therefore
assume that $N$ scalars can only get vevs of the order of the weak or SUSY
breaking scales, i.e. a few TeV or less; in this case the bound of
ref.\cite{10} holds.

The existence of an upper bound on the mass of an exotic fermion does not yet
mean that we can test this model decisively, however. To begin with, present
\epem\ colliders do not have sufficient energy to produce even the lightest
exotics for all combinations of parameters. This will change as soon as the
next generation of linear \epem\ colliders goes into operation. However, even
here the neutral exotics might be impossible to find. We will argue that the
lightest neutral exotic lepton is likely to be stable. In this case it will
only give rise to an observable signal at colliders if it is produced in
association with a heavier exotic. Unfortunately {\em no} upper bound on the
masses of these heavier states can as yet be given.

In this seemingly hopeless situation cosmological arguments come to the
rescue. It turns out that the upper bound on the mass of the lightest exotic
{\em decreases} as the experimental lower bounds on the masses of the heavier
exotics increase. Moreover, in this situation the light exotics are forced to
have smaller and smaller $SU(2)$ doublet components, which suppresses their
couplings to gauge and Higgs bosons. Since the light exotics are stable, some
fraction of such particles produced in the very early universe is still around
today \cite{11}. This fraction, and hence the contributions of such exotic Big
Bang relics to the present matter density, is (approximately) inversely
proportional to the annihilation cross section of the exotics. The crucial
observation is that light, singlet--dominated exotics will have small
annihilation cross sections by virtue of their suppressed couplings, and hence
large relic densities. It is therefore clear that at some point the lower
experimental bound on the masses of the {\em heavier} exotic leptons can force
the cosmological relic density of the {\em light} exotic to become
unacceptably large. We find that, before this happens, light neutral exotics
will become accessible to Higgs boson decays, leading to a large invisible
branching ratio for at least one light Higgs boson. It is the purpose of the
present paper to study these connections quantitatively. Of course, the
heavier exotics might well be discovered at rather low masses, in which case
the relic density constraint could be fulfilled easily. In this optimistic
scenario one can study the properties of the new fermions in order to further
test the model. We wish to emphasize here that at some future time, the
combination of collider searches and the relic density constraint will test
the model {\em decisively}, independent of the values of the (many) free
parameters. Unfortunately we find that a 500 GeV \epem\ collider is not quite
sufficient to probe the entire parameter space; one will have to push the
energy to 1 or even 1.5 TeV in order to close the last loophole. Nevertheless
we find it remarkable that a decisive test is possible at all; after all, most
experimental bounds can be evaded by simply increasing the ``new physics"
scale.

The remainder of this paper is organized as follows. In Sec.~2 we describe
those parts of the model that are of relevance to us, i.e. the exotic leptons
and Higgs bosons; we also give the couplings of the (neutral) exotics to gauge
and Higgs bosons. In Sec.~3 the cross sections for the production of neutral
exotics at \epem\ colliders are listed. In Sec.~4 the calculation of the
cosmological relic density of the light neutral exotics is discussed; some
care must be taken here, due to the prominent role played by $s-$channel
exchange diagrams, as well as the presence of a second light exotic, which can
suppress the relic density by co--annihilation. Sec.~5 contains our numerical
results, and Sec.~6 is devoted to a brief summary.

\setcounter{footnote}{0}
\section*{2) The Model}
For defineteness we will work in the framework of the minimal rank 5 subgroup
of $E(6)$ \cite{1}, which does not require the introduction of an intermediate
scale between the GUT and weak scales. However, our analysis would go through
with only minor modifications in models with a gauge group of rank 6 at the
weak scale, as well as in models where some $\tilde{\nu}_R$ field has a vev
around $10^{10}$ GeV or more. As already stated in the Introduction, the only
important assumption we have to make is that none of the scalar $N$ fields
gets a vev much above $10^9$ GeV.

In $E(6)$ models each {\bf 27}--dimensional representation contains one $N$
superfield, which is an SM singlet, as well as the exotic $SU(2)$ doublet
superfields $H$ and $\bar{H}$, whose scalar components have just the right
quantum numbers to serve as the Higgs bosons of the MSSM, i.e. to provide
masses for the $W$ and $Z$ bosons. At least one of the $N$ scalars also has to
get a vev in order to give a sufficiently large mass to the single new $Z'$
boson of the rank--5 model. Using rotations in generation space, we can always
work in a basis where only one $H^0$, one $\bar{H^0}$ and one $N$ field have
non--vanishing vev; following the notation of ref.\cite{9} we call these true
Higgs fields \hth, \hbt\ and \nt. Their fermionic superpartners then mix with
the superpartners of the three neutral gauge bosons to form the six neutralino
states of this model \cite{9}.

Here we are interested in the first two generations of fermionic $H, \
\bar{H}$ and $N$ fields. They obtain their masses from the superpotential
\be \label{e1}
W_{\rm lep} = \sum_{i,j,k=1}^3 \la_{ijk} H_i \bar{H}_j N_k.
\ee
In order to give masses to all charged exotic leptons\footnote{We call these
fields ``leptons" merely in order to indicate that they do not have strong
interactions; they need not carry the same lepton number as the charged
leptons and neutrinos of the SM}, we at least have to allow those couplings
where exactly one of the indices $i,j,k$ in eq.(\ref{e1}) equals 3, the other
two being either 1 or 2. By further rotations between fields of the first two
generations we can define $\la_{123} = \la_{213} = 0$, without loss of
generality; the contribution of these couplings to the mass matrix of the
charged exotic fermions is then diagonal, with
\ben \label{e2} \beq
\mlop &= \la_{113} x; \label{e2a} \\
\mltp &= \la_{223} x, \label{e2b}
\eeq \een
where we have introduced $x \equiv \langle \nt \rangle$.

In the basis $(\wt{H_1^0}, \wt{\bar{H^0_1}}, \wt{N_1}, \wt{H_2^0},
\wt{\bar{H^0_2}}, \wt{N_2})$, the contribution of these couplings to the mass
matrix of the neutral exotic leptons reads:
\be \label{e3}
{\cal M}_{L^0} = \mbox{$
\left( \begin{array}{cccccc}
0 & \mlop & \la_{131} \vb & 0 & 0 & \la_{132} \vb \\
\mlop & 0 & \la_{311} v & 0 & 0 & \la_{312} v \\
\la_{131} \vb & \la_{311} v & 0 & \la_{231} \vb & \la_{321} v & 0 \\
0 & 0 & \la_{231} \vb & 0 & \mltp & \la_{232} \vb \\
0 & 0 & \la_{321} v & \mltp & 0 & \la_{322} v \\
\la_{132} \vb & \la_{312} v & 0 & \la_{232} \vb & \la_{322} v & 0
\end{array} \right) $},
\ee
with $v \equiv \langle \hth \rangle, \ \vb \equiv \langle \hbt \rangle$. The
neutral exotic leptons are Majorana fermions; the mass matrix (\ref{e3}) is
therefore symmetric.

The coupling $\la_{333}$ in eq.(\ref{e1}) must also be non--zero, in order to
avoid the presence of a dangerous axion (see below). Terms where none of the
three subscripts equals 3 do not contribute to the mass matrices. Finally,
there could be contributions to the superpotential (\ref{e1}) where only one
of the three indices is not equal to 3. In this case the charged exotic
leptons would mix with the charginos, and the neutral exotic leptons would mix
with the neutralinos. However, if such couplings are present, the rotatations
in field space that define the basis where only \hth, \hbt\ and \nt\ have
non--zero vevs would depend on the renormalization scale. Worse, all $H_i$ and
$\bar{H}_i$ would then couple to SM quarks and leptons at least at the one
loop level, which could lead to dangerous flavour changing neutral currents
(FCNC) \cite{12}. We therefore forbid these terms. Fortunately, this is not
only technically natural, but even follows automatically if one requires all
potentially dangerous terms in the low energy superpotential (which can lead to
fast proton decay, large neutrino masses, or tree--level FCNC) to be forbidden
by a single discrete $Z_2$ symmetry \cite{13}.

This further restriction of the allowed terms in eq.(\ref{e1}) greatly
simplifies the calculation, and leads to a much more appealing model, but has
no significant impact on the properties that are of interest to us. This can
be seen from the fact that even if all terms in eq.(\ref{e1}) were present,
in the limit $v,\vb \rightarrow 0$ the $12 \times 12$ neutral fermion mass
matrix would have two zero eigenvalues, corresponding to SM singlet fermions
$\wt{N}_1, \ \wt{N}_2$. Since all couplings in eq.(\ref{e1}) are required to
be of order 1 or less \cite{6,10}, this observation implies that even in the
realistic situation with non--vanishing $v$ and \vb, there will be two neutral
Majorana fermions whose masses are roughly of order $M_Z$ or less. Moreover,
the upper bound on the masses of these neutral fermions will decrease as the
(experimental lower bounds on the) masses of the charged exotic fermions is
increased. We repeat, these crucial properties of the model follow from
eq.(\ref{e1}) without further assumptions. For the remainder of our analysis
we will concentrate on the simpler case where the mass matrix (\ref{e3}) is
decoupled from the neutralino mass matrix, just to avoid needless
complications.

A little calculation shows that the determinant of the matrix (\ref{e3}) is
proportional to $(v \vb x)^2$; this means that all three vevs have to be
non--zero if all eigenvalues are to be non--vanishing. It is also quite easy
to see that in the limit $m_{L^\pm_{1,2}} \gg v,\vb$, and with all couplings
${\cal O}(1)$ or less, there will be four large eigenvalues, approximately
equal to $\pm \mlop$ and $\pm \mltp$. The other two eigenvalues are \cite{9}
then of order $\la^2 v \vb / m_{L^\pm_{1,2}}$. This is why the upper bound on
the smaller eigenvalues decreases with increasing mass of the charged exotics,
as stated above. Indeed, in ref.\cite{10} it was shown that the mass \ml\ of
the lightest neutral exotic lepton is maximized if all entries of the mass
matrix (\ref{e3}) are of the same order (or zero). Numerically one has $\ml
\leq 110$ GeV if $m_{L^\pm_{1,2}} \geq 45$ GeV.

In general the matrix (\ref{e3}) has to be diagonalized numerically. The
resulting eigenstates $L_i^0$ are given by
\be \label{e4}
L_i^0 = \sum_{j=1}^6 U_{ij} \wt{N}_j,
\ee
where we have introduced the 6--component vector $\wt{N} \equiv (\wt{H_1^0},
\wt{\bar{H^0_1}}, \wt{N_1}, \wt{H_2^0}, \wt{\bar{H^0_2}}, \wt{N_2})$, and $U$
is an othogonal\footnote{We assume that all couplings in eq.(\ref{e1}) are
real.} matrix chosen such that
\be \label{e5}
U {\cal M}_{L^0} U^T = {\rm diag} (m_{L_i^0}), \ \ i=1, \dots, 6.
\ee
The couplings of the charged and neutral exotic mass eigenstates to the
standard $Z$ boson are then given by the Lagrangean \cite{14}
\be \label{e6}
{\cal L}_{ZLL} = \frac {g} {2 \ctw} Z_\mu \left[ \sum_{a=1}^2
\overline{L_a^\pm} \gamma^\mu L_a^\pm (1 - 2 \sin^2 \theta_W )
+ \sum_{i,j=1}^6 \overline{L_i^0} \gamma^\mu \gamma_5 L_j^0 t_{ij} \right],
\ee
where $g$ is the $SU(2)$ gauge coupling, $\theta_W$ the weak mixing angle, and
\be \label{e7}
t_{ij} = \frac{1}{2} \left( U_{i2} U_{j2} + U_{i5} U_{j5}
 - U_{i1} U_{j1} - U_{i4} U_{j4} \right).
\ee

A few comments are in order here. First, when writing
eqs.(\ref{e5}),(\ref{e6}) we have allowed the masses of the neutral exotic
leptons to have either sign. We could also insist that all $m_{L_i^0} \geq 0$
by inserting appropriate factors of $i$ in the matrix $U$, which would then no
longer be orthogonal (but would still be unitary, of course). In this case the
diagonal couplings of the $Z$ boson to neutral exotics would still be purely
axial vector, but the off--diagonal couplings would become purely vector if
the two corresponding eigenvalues have opposite signs. Secondly, recall that
the $L_i^0$ are (4--component) Majorana spinors. This means that their
couplings appearing in Feynman diagrams are twice as large as those in the
Lagrangean (\ref{e6}). Finally, following ref.\cite{9}, and contrary to the
usual MSSM notation, we have defined $H_{1,2,3}$ to have hypercharge $+ 1/2$,
while the hypercharge of $\bar{H}_{1,2,3}$ is $-1/2$; this explains the
overall sign of the coupling $t_{ij}$ in eq.(\ref{e7}).

In our calculation of cosmological relic densities we also have to specify the
Higgs sector of the model. Recall that we are working in a basis where only
\hth, \hbt\ and \nt\ have non--zero vevs. The relevant part of the Higgs
potential is then given by \cite{15}
\beq \label{e8}
V_{\rm Higgs} &= m^2_{H_3} | \hth |^2 + m^2_{\bar{H}_3} | \hbt |^2
+ m^2_{N_3} | \nt |^2 + (\la_{333} A_{333} \hth \hbt \nt + h.c.) \nonumber \\
&+ \la^2_{333} \left[ | \hth \hbt |^2 + | \nt |^2 \left( | \hth |^2 +
| \hbt |^2 \right) \right] \\
&+ \frac {g^2} {8 \cos^2 \theta_W} \left( | \hth |^2 - | \hbt |^2 \right)^2
+ \frac {g'^2}{72} \left( 5 | \nt |^2 - 4 | \hbt |^2 - | \hth |^2 \right)^2,
\nonumber \eeq
where $g'$ is the $U(1)$ gauge coupling. We will assume the Higgs mass
parameters as well as the soft breaking parameter $A_{333}$ to be real. The
vevs can then all chosen to be real \cite{16}, and the mass matrices for the
real (scalar) and imaginary (pseudoscalar) parts of the neutral Higgs fields
decouple. The former is given by \cite{9}
\be \label{e9}
{\cal M}^2_H = \mbox{$
\left( \begin{array}{ccc}
\left( \frac{g^2}{2} + \frac {25}{18} g'^2 \right) v^2
- \ta \frac {\bar{v} x}{v} &
\left(2 \la^2_{333} -\frac{g^2}{2} -\frac{5}{18} g'^2 \right) v \vb + \ta x &
\left( 2 \la^2_{333} - \frac{10}{9} g'^2 \right) vx + \ta \vb \\
\left(2 \la^2_{333} -\frac{g^2}{2} -\frac{5}{18} g'^2 \right) v \vb + \ta x &
\left( \frac{g^2}{2} + \frac {5}{9} g'^2 \right) \vb^2
- \ta \frac {v x}{\bar{v}} &
\left( 2 \la^2_{333} - \frac{5}{18} g'^2 \right) \vb x + \ta v \\
\left( 2 \la^2_{333} - \frac{10}{9} g'^2 \right) vx + \ta \vb &
\left( 2 \la^2_{333} - \frac{5}{18} g'^2 \right) \vb x + \ta v &
\frac{25}{18} g'^2 x^2 - \ta \frac {v \bar{v}}{x}
\end{array} \right) $},
\ee
with $\ta \equiv \la_{333} A_{333}$. In writing eq.(\ref{e9}) we have used the
requirement that the first derivatives of the potential (\ref{e8}) with
respect to the fields should vanish in the minimum. Notice that the smallest
eigenvalue $m^2_{h_1^0}$ of the matrix (\ref{e9}) is again only of order
$v^2 + \vb^2$, not of order of the $Z'$ boson mass or the SUSY breaking scale
\cite{17}:
\be \label{e10}
m^2_{h_1^0} \leq M_Z^2 \left[ \cos^2 (2 \beta) + \frac {2 \la^2_{333} \cos^2
\theta_W} {g^2} \sin^2(2 \beta) + \frac {\sin^2 \theta_W} {9} \left( 4
\sin^2 \beta + \cos^2 \beta \right)^2 \right],
\ee
where $\tanb \equiv v / \vb \ (> 1)$. Eqs.(\ref{e9}) and (\ref{e10}) only hold
at tree level. There are sizable radiative corrections from top and stop
loops, as well as possibly from loops involving exotic (s)quarks and
(s)leptons \cite{18}. However, for our purpose their effect can largely be
mimicked by allowing $\la_{333}$ at the weak scale to be quite large. The
reason is that these corrections tend to increase the mass of the lightest
scalar Higgs boson; eq.(\ref{e10}) shows that increasing $\la_{333}$ has the
same effect. Moreover, the radiative corrections are only important if they
involve large Yukawa couplings, e.g. that of the top quark. However, the
introduction of these large couplings also {\em reduces} the upper bound on
$\la_{333}$ which follows from the requirement that the model remains weakly
interacting up to the GUT scale. The final upper bound on $m_{h_1^0}$ is
therefore not much changed by these corrections.

Since the model contains two massive neutral gauge bosons, only one physical
pseudoscalar $A$ survives. It can be written as \cite{9}
\beq \label{e11}
A &= \frac {1} {\sqrt{ \bar{v}/v + v/\bar{v} + v \bar{v} / x^2 }}
\frac{1} {\sqrt{2}} \Im \left( \frac{\vb}{v} \hth + \frac{v}{\vb} \hbt +
\frac {v \vb}{x^2} \nt \right) \nonumber \\
&\equiv \frac{1}{\sqrt{2}} \sum_{j=1}^{3} P_j \Im \left( H_j \right),
\eeq
where we have introduced the complex 3--component vector $H \equiv \left( \hth,
\hbt, \nt \right)$, and the symbol $\Im$ denotes the imaginary part. The mass
of the physical pseudoscalar is given by \cite{9} \be \label{e12}
m_A^2 = -\la_{333} A_{333} x \left( \frac{\vb}{v} + \frac{v}{\vb} +
\frac{v \vb}{x^2} \right).
\ee
Notice that this state would be massless if $\la_{333}=0$, as mentioned
earlier. There is also a physical charged Higgs boson, with mass \cite{9}
\be \label{ech}
m^2_{H^\pm} = M_W^2 ( 1 - \frac {2 \la^2_{333}} {g^2} )
+ m_A^2 + \la_{333} A_{333} \frac {v \vb} {x}.
\ee
Note that this Higgs boson can be lighter than the $W$ boson if $m_A$ is small
and $\la_{333}$ is sizable.

In general the mass matrix (\ref{e9}) is most easily diagonalized numerically.
Since it is real and symmetric, the diagonalization can be achieved by an
orthogonal matrix $S$:
\be \label{e13}
S {\cal M}^2_H S^T = {\rm diag}(m^2_{h_1^0}, m^2_{h_2^0}, m^2_{h_3^0} ).
\ee
The physical scalars $h_i^0$ are then given by
\be \label{e14}
h_i^0 = \frac{1}{\sqrt{2}} \sum_{j=1}^3 S_{ij} \Re \left( H_j \right).
\ee
The matrix $S$ and the vector $P$ introduced in eq.(\ref{e11}) determine the
couplings of the physical Higgs bosons to SM particles. We need the
$Z-Z-h^0_i$ couplings in order to interpret bounds on the Higgs sector from
searches at LEP; they are given by
\be \label{e15}
g_{ZZh_i^0} = \frac {g M_Z} {\cos \theta_W} \left( S_{i1} \sinb + S_{i2}
\cosb \right).
\ee
Bounds on $Z \to h_i^0 A$ decays involve the couplings
\be \label{e16}
g_{ZAh_i^0} = \frac{g} {2 \cos \! \theta_W} \left( S_{i1} P_1 -
S_{i2} P_2 \right).
\ee
In Sec.~4 we will also need the couplings of the Higgs bosons to the quarks
and leptons of the SM. We write the corresponding Lagrangean as
\be \label{e17}
{\cal L}_{Hff} = \frac{g} {\ctw} \bar{f} \left[ c^{(f)} (-i \gamma_5) A +
\sum_{i=1}^3 d_i^{(f)} h_i^0 \right] f,
\ee
with
\ben \label{e18} \beq
c^{(f)} &= -\frac {m_f}{2 M_Z} \cdot \left\{ \begin{array}{ll}
P_1/\sinb, \ \ \ & f=u,c,t \\
P_2/\cosb, \ \ \ & f=e,\mu,\tau,d,s,b
\end{array} \right. \label{e18a} \\
d_i^{(f)} &= -\frac {m_f}{2 M_Z} \cdot \left\{ \begin{array}{ll}
S_{i1}/\sinb, \ \ & f=u,c,t \\
S_{i2}/\cosb, \ \ & f=e,\mu,\tau,d,s,b
\end{array} \right. \label{e18b}
\eeq \een

Finally, we will need the couplings of the physical Higgs bosons to the
neutral exotic leptons. They can most easily be expressed in terms of the
couplings $\la'_{ijk}$ between the lepton and Higgs current eigenstates:
\be \label{e19}
{\cal L}_{\wt{N} \wt{N} H} = - \frac{1}{4} \sum_{i,j=1}^6 \sum_{k=1}^3
\la'_{ijk} \overline{\wt{N}}_i (1 - \gamma_5) \wt{N}_j H^0_k + h.c.,
\ee
where $\wt{N}$ is the 6--component vector of Majorana spinors introduced in
eq.(\ref{e4}), and $H$ the 3--component vector of complex neutral Higgs fields
defined in eq.(\ref{e11}). The couplings $\la'_{ijk}$ are determined by the
superpotential (\ref{e1}):
\beq \label{e20}
\la'_{123} &= \la_{113}, \ \la'_{132} = \la_{131}, \
\la'_{153} = \la_{123}, \ \la'_{162} = \la_{132}, \
\la'_{231} = \la_{311}, \ \la'_{243} = \la_{213}, \nonumber \\
\la'_{261} &= \la_{312}, \ \la'_{342} = \la_{231}, \
\la'_{351} = \la_{321}, \ \la'_{453} = \la_{223}, \
\la'_{462} = \la_{232}, \ \la'_{561} = \la_{322};
\eeq
this has to be symmetrized in the first two indices ($\la'_{ijk} = \la'_{jik}
$), and all other $\la'$ couplings vanish. Recall that without loss of
generality we can work in a basis where the mass matrix for the charged exotic
leptons is diagonal, i.e. $\la_{123} = \la_{213} = 0$; this implies
$\la'_{153} = \la'_{243} = 0$ in this basis. The interaction between lepton
and Higgs mass states can then be written as
\beq \label{e21}
{\cal L}_{LLH} &= - \frac {1} {2 \sqrt{2}} \sum_{i,j=1}^6 \sum_{k=1}^3
\la'_{ijk} \sum_{l=1}^3 S_{lk} h^0_l \sum_{m,n=1}^6 U_{mi} U_{nj}
\overline{L^0_m} L^0_n \nonumber \\
&+ \frac {i} {2 \sqrt{2}} \sum_{i,j=1}^6 \sum_{k=1}^3 \la'_{ijk}
P_k A \sum_{m,n=1}^6 U_{mi} U_{nj} \overline{L^0_m} \gamma_5 L^0_n,
\eeq
where the orthogonal matrices $U$ and $S$ have been defined in eqs.(\ref{e5})
and (\ref{e13}), respectively, and the eigenvector $P$ of the pseudoscalar
mass matrix is given in eq.(\ref{e11}). Recall that the $L^0_m$ are Majorana
fermions, and the physical Higgs bosons $h^0_l$ and $A$ are described by real
fields.
\section*{3) Constraints from \epem\ Colliders}
In this section we discuss how new physics searches at existing and future
\epem\ colliders can constrain the class of $E(6)$ models we are considering.
At present the most stringent constraints on both the exotic leptons and the
Higgs sector come from LEP. To begin with, the failure to observe the charged
exotic leptons immediately implies that their masses must exceed 45 GeV, since
eq.(\ref{e6}) shows that they always couple with essentially full gauge
strength to the $Z$ boson.

The decay of the $Z$ into two neutral exotic leptons needs a somewhat more
detailed discussion. The heavier $L_i^0$ states can always decay into $L_1^0$
plus a real or virtual $Z$ or Higgs boson. However, the lightest state $L_1^0$
can only decay into SM particles or the right--handed neutrino state $\nu_R$,
or their superpartners. Such decays could be due to terms in the
superpotential of the type $\bar{H}_{1,2} l_L e_R, \ \bar{H}_{1,2}q_L d_R$ or
$H_{1,2} q_L u_R$, where $q_L$ and $l_L$ are quark and lepton $SU(2)$ doublet
superfields and $e_R, \ d_R$ and $u_R$ the corresponding singlets. However,
couplings of this kind will in general generate tree--level FCNC; these terms
are therefore severely constrained \cite{19}, which is why they are often
completely forbidden in (potentially) realistic models, e.g. by means of a
discrete symmetry \cite{13}. The model in general also allows couplings of the
kind $H_{1,2} l_L \nu_R$. Such couplings can lead to flavor changing processes
in the lepton sector ($\mu \to e \gamma, \ \tau \to \mu \gamma$, etc), but
only at one--loop level; they are therefore somewhat less tightly constrained
\cite{19}. If any of these couplings exists, the exotic leptons become odd
under $R$ parity. This means that the decay product of $L_1^0$ must contain a
superparticle. Since \lone\ is quite light, we can expect that only final
states containing the lightest sparticle, which in almost all cases is the
lightest neutralino $\wt{Z}_1$, can occur. The least tightly constrained of
the above couplings, $H_{1,2} l_L \nu_R$, will then lead to an invisible final
state, if we assume that $R$ parity is conserved. Of course, it is not at all
clear that such decays are kinematically allowed, since $\wt{Z}_1$ could very
well be heavier than $L_1^0$. Indeed, in view of the fact that present bounds
\cite{3,8} require the $Z'$ boson mass to exceed several hundred GeV, we need
a rather high value of the SUSY breaking scale in this model \cite{5,6}. We
therefore conclude that $L_1^0$ most probably is either absolutely stable, or
decays invisibly; in particular, this is always true if tree--level FCNC are
suppressed by a simple symmetry and $R$ parity is conserved. In this case
$L_1^0$ cannot be detected directly by collider experiments. Since we want to
devise a test of the model that works even under the least favorable
circumstances we will assume that $L_1^0$ is indeed invisible.

The partial width of $Z \to L_i^0 L_j^0$ decays is given by \cite{14}:
\beq \label{e22}
\Gamma ( Z \to L_i^0 L_j^0 ) &= \frac { |\vec{k}| } {24 \pi M_Z^2}
(2 - \delta_{ij}) \left( \frac {g t_{ij}} {\ctw} \right)^2 \nonumber \\
& \cdot \left[ 2M_Z^2 - m_i^2 - m_j^2 - 6 m_i m_j - \frac { \left( m_i^2 -
m_j^2 \right)^2 } {M_Z^2} \right],
\eeq
where we have used the shorthand notation $m_i \equiv m_{L_i^0}$. The coupling
$t_{ij}$ has been defined in eq.(\ref{e7}), and the $L_i^0$ 3--momentum in the
$Z$ rest frame $|\vec{k}|$ is given by
\be \label{e23}
|\vec{k}| = \frac {1} {2 M_Z} \sqrt{ \left( M_Z^2 - m_i^2 - m_j^2 \right)^2
- 4 m_i^2 m_j^2}.
\ee
Recall that in our convention the neutral lepton masses, and hence the
bi--linear term in eq.(\ref{e22}), can have either sign.

We have argued above that the combination $i=j=1$ only contributes to the
invisible width of the $Z$ boson. However, even though the heavier neutral
leptons will almost always decay inside the detector, their production might
not lead to an experimentally observable final state if the mass difference
to the lightest lepton is too small. We have quite conservatively required
$|m_i| - |m_1| \geq 3$ GeV for an ``experimentally visible" $L_i^0$. The
signature for the production of these visible states is quite similar to that
for the production of the heavier neutralino states of the MSSM. Searches
\cite{20} for such neutralinos have so far not been sucessful; the bound on
the resulting branching ratio is of the order of (a few times) $10^{-5}$, the
precise value depending on the masses of the particles involved. We have
therefore imposed the upper bound
\be \label{e24}
\Gamma (Z \to {\rm visible \ exotics}) < 0.13 \ {\rm MeV},
\ee
which corresponds to a bound on the branching ratio of about $5 \cdot
10^{-5}$.

The best bound on non--SM contributions to invisible $Z$ decays comes from the
measurement of $\Gamma_{\rm inv}(Z) / \Gamma(Z \to l^+ l^-)$; this can be
interpreted \cite{21} as a measurement of the number of SM neutrino species,
$N_\nu = 2.988 \pm 0.023$. Taking into account that the model predicts $N_\nu
\geq 3$, this translates into the bound
\be \label{e25}
\Delta \Gamma_{\rm inv} \leq 6.7 \ {\rm MeV}
\ee
at 95\% confidence level. Note that ``invisible" includes everything that is
not counted in any explicitly reconstructed final state; in particular, $L_i^0
L_j^0$ final states where neither of the two leptons is at least 3 GeV heavier
than $L_1^0$ are included here.

At future high energy \epem\ colliders the neutral exotic leptons can be
produced via the exchange of a virtual $Z$ boson. The cross section is given
by
\beq \label{e26}
\sigma( \epem \to L_i^0 L_j^0) &= \frac {|\vec{k}|} {4 \pi \rs} ( 2 -
\delta_{ij} ) \left( \frac {g} {\ctw} \right)^4 \frac { \left( v_e^2 + a_e^2
\right) t_{ij}^2} { \left( s - M_Z^2 \right)^2 + M_Z^2 \Gamma_Z^2 }
\nonumber \\
& \cdot \left( E_i E_j - m_i m_j + \frac {1}{3} |\vec{k}|^2 \right),
\eeq
with $a_e = -1/4$ and $v_e = 1/4 - \sin^2 \theta_W$. $|\vec{k}|$ is given by
eq.(\ref{e23}) with $M_Z$ replaced by the center--of--mass energy \rs, and
$E_i = \sqrt{ m_i^2 + |\vec{k}|^2 }$. In principle this cross section also
receives contributions from $Z'$ boson exchange. However, if these
contributions are sizable, $Z'$ exchange will also lead to observable effects
in the pair production of SM fermions. Future \epem\ colliders will therefore
be able \cite{2} to find evidence for a $Z'$ boson with mass up to several
times \rs. The discovery of a $Z'$ signal would not only (obviously) rule out
the SM, but would also allow significant new tests of the type of model we are
studying here. In order to be conservative we therefore always assume that the
contributions from $Z'$ exchange are too small to be detectable. Finally, we
will always assume that future \epem\ colliders will probe for the existence
of charged exotic leptons almost up to the kinematical limit, since they have
full strength couplings to both the $Z$ boson and the photon.

The Higgs sector of the model is also constrained by unsuccessful new physics
searches at LEP. We have used the results of the ALEPH collaboration
\cite{22}, which has published bounds on the $ZZh_i^0$ and $ZAh_i^0$ couplings
of light Higgs bosons. Their numerical bounds can be approximated by:
\ben \label{e27} \beq
\left( S_{i1} \sin \! \beta + S_{i2} \cos \! \beta \right)^2 & \leq
\left\{ \mbox{$ \begin{array} {ll}
 0.025, & m_{h_i^0} \leq 10 \ {\rm GeV} \\
 0.005 m_{h_i^0} - 0.025, & 10 \ {\rm GeV} \leq m_{h_i^0} \leq 20 \ {\rm GeV}
\\
 0.0175 m_{h_i^0} - 0.275,& 20 \ {\rm GeV} \leq m_{h_i^0} \leq 30 \ {\rm GeV}
\\
 0.025 m_{h_i^0} - 0.5,& 30 \ {\rm GeV} \leq m_{h_i^0} \leq 60 \ {\rm GeV}
\end{array} $} \right. \label{e27a} \\
\left( S_{i1} P_1 - S_{i2} P_2 \right)^2 & \leq \left\{ \mbox{$
\begin{array} {ll}
0.1, & m_{h_i^0} + m_A \leq 81 \ {\rm GeV} \\
0.1 \left( m_{h_i^0} + m_A - 81 \ {\rm GeV} \right) + 0.1, & m_{h_i^0} + m_A
\leq M_Z
\end{array} $} \right. \label{e27b}
\eeq \een
where all masses are in GeV.

As emphasized in Sec.~2, at least one of the three neutral scalar Higgs bosons
of the model must have mass below 150 GeV or so; moreover, this boson will
couple to the $Z$ with full strength if the bound (\ref{e10}) on its mass is
saturated. For some combinations of parameters the lightest Higgs scalar will
have a very weak coupling to the $Z$; however, in this case the
next--to--lightest Higgs scalar will have unsuppressed coupling to the $Z$,
and its mass will also satisfy the bound (\ref{e10}). In fact, the scalar
Higgs sector of our model is quite similar to that of the MSSM with additional
Higgs singlet superfield, where it has been shown \cite{23} that an \epem\
collider with $\rs \geq 300$ GeV has to detect at least one neutral scalar
Higgs boson. Unfortunately this may not be sufficient to allow a significant
test of the model, since this Higgs boson might look very similar to the
single Higgs boson of the SM. Even if several Higgs bosons are discovered, it
might be quite difficult to distinguish between the $E(6)$ model were are
discussing here and the MSSM, since large $x$ means that one of the three
neutral Higgs scalars is quite heavy and essentially a pure singlet; the
singlet components of the other two scalar Higgs bosons and the single
pseudoscalar Higgs boson are then very small. We therefore try to avoid making
assumptions about searches for neutral Higgs bosons at future colliders as
much as possible. There is one exception, however: The light Higgs boson(s) of
the model might have large branching ratios into neutral exotic leptons, which
could lead to a large invisible branching ratio for these Higgs particles.
This would be fairly distinctive; in the MSSM a large invisible branching
ratio of a Higgs boson would imply that the light chargino will be discovered
at LEP2 \cite{24}. In contrast, we will always require that the charged exotic
leptons are too heavy to be produced at the \epem\ collider under
consideration.

The partial widths for Higgs decays into exotic lepton can be written as
\be \label{e28}
\Gamma (H \to L_i^0 L_j^0) = \frac {\la^2} {4 \pi m_H^3} (2 - \delta_{ij})
\sqrt{ m_H^2 - (m_i+m_j)^2 } \sqrt{ m_H^2 - (m_i-m_j)^2 }
 \cdot \left[ m_H^2 - \left( m_i \pm m_j \right)^2 \right].
\ee
As usual, the lepton masses $m_{i,j}$ can have either sign here. The ``$+$"
(``$-$") sign in the last term of eq.(\ref{e28}) applies if $H$ is scalar
(pseudoscalar), and \la\ is the relevant $H L_i^0 L_j^0$ coupling in the
Lagrangean (\ref{e21}); recall that a factor of $2 \la$ appears in the
relevant Feynman rule, due to the Majorana nature of $L_i^0$. When computing
the total decay widths of the neutral Higgs bosons of our model, we include
decays into SM fermion pairs as well as (for the scalar Higgs bosons) $W^+W^-$
and $ZZ$ final states, where we allow one (but not both) of the gauge bosons
to be off--shell. Expressions for the relevant partial widths can, e.g., be
found in the recent review article \cite{24}; of course, we have to use the
couplings listed in eqs.(\ref{e15})--(\ref{e18}) here. Note that the quark
masses in eqs.(\ref{e18}) are meant to be running masses, taken at scale
$m_H$; we also include the leading non--logarithmic QCD corrections \cite{25}
(which remain finite in the limit $m_q/m_H \to 0$) to the $c \bar{c}$ and $b
\bar{b}$ partial widths. An accurate estimate of the decay widths of the light
Higgs bosons of the model is necessary not only for the evaluation of
branching ratios for exotic Higgs decays, but also for the calculation of the
cosmological relic density, to which we turn next.

\section*{4) The Cosmological Relic Density}

We argued in the previous section that the exotic leptons will have odd $R$
parity if they have any Yukawa interactions with ordinary matter. The lightest
exotic lepton \lone\ will then be absolutely stable if it is lighter than
the lightest neutralino $\wt{Z}_1$. The only way out would be to introduce
$R$ parity breaking interactions in the superpotential. However, in this case
the model, although very complicated, would no longer be able to explain the
existence of cold dark matter (CDM); this seems to be required \cite{25a} for
a successful fit of all data on large scale structure in the universe. The
fact that $\wt{Z}_1$ makes an excellent CDM candidate is often viewed as one
of the strengths of the MSSM; a significantly more complicated model that
gives up this advantage does not look very appealing to us. We will therefore
require $R$ parity to be conserved.

Unfortunately the usual approximation \cite{11} for the calculation of the
relic density is often not reliable in this model. Two of the three cases
discussed in ref.\cite{26} as examples where the usual method is not
applicable can occur here, possibly simultaneously: Narrow $s-$channel (Higgs)
poles play a very important role in the calculation of the annihilation cross
section; and the next--to--lightest neutral exotic \ltwo\ can be quite close
in mass to \lone, in which case \lone\ltwo\ co--annihilation as well as
\ltwo\ltwo\ annihilation have to be considered as well.

We have followed ref.\cite{26} in our computation of the relic density. For
completeness we list, but do not derive, the relevant expressions here. The
calculation proceeds in two steps. First one has to determine the freeze--out
temperature $T_f$, where the relic particles fall out of thermal equilibrium.
It is given by
\be \label{e29}
x_f = \ln \frac {0.038 g_{\rm eff} M_P |m_1| \sigav (x_f)}
{\sqrt{ g_* x_f}},
\ee
where we have introduced $x_f \equiv |m_1|/T_f$, $m_1$ being the mass of the
lightest neutral exotic. $M_P = 1.22 \cdot 10^{19}$ GeV is the Planck mass,
and $g_*$ is the number of degrees of freedom that are in thermal equilibrium
at temperature $T_f$; we have for simplicity used a fixed $\sqrt{g_*}=9$,
which introduces a negligible error in our calculation. Eq.(\ref{e29}) has to
be solved iteratively, since the r.h.s. depends on $x_f$ both explicitly and
via the thermal average \sigav, defined as
\be \label{e30}
\sigav (x) = \frac {x^{3/2}} {2 \sqrt{\pi}} \int_0^\infty dv v^2 e^{-v^2 x/4}
\sigma_{\rm eff} v,
\ee
where $v$ is the relative velocity of the two annihilating particles in their
center--of--mass frame. In the presence of co--annihilation, the effective
annihilation cross section $\sigma_{\rm eff}$ is given by
\be \label{e31}
\sigma_{\rm eff} (x) = \frac {4} {g_{\rm eff}^2} \left[ \sigma_{11} + 2
\sigma_{12} ( 1 + \Delta)^{3/2} e^{-x \Delta} + \sigma_{22} (1 + \Delta)^3
e^{-2x \Delta} \right],
\ee
where $\sigma_{ij} = \sigma(L_i^0 L_j^0 \to$ anything), and we have introduced
\ben \label{e32} \beq
\Delta &= \left| \frac {m_2} {m_1} \right| - 1; \label{e32a} \\
g_{\rm eff} &= 2 \left[ 1 + (1 + \Delta)^{3/2} e^{-x \Delta} \right].
\label{e32b} \eeq \een
$g_{\rm eff}$ is the effective number of degrees of freedom of CDM particles
($g=2$ for a single Majorana fermion). Numerically, $x_f \simeq 20$ or so for
models leading to an acceptable relic density.

Once $x_f$ has been determined, the relic density is given by
\be \label{e33}
\omeh = \frac { 1.07 \cdot 10^9 \ {\rm GeV}^{-1} }
{J(x_f) \sqrt{g_*} M_P },
\ee
where the annihilation integral $J$ is defined as
\be \label{e34}
J(x_f) = \int_{x_f}^\infty \frac {\sigav (x) } {x^2} dx.
\ee
As usual, we have expressed the relic density $\Omega$ in units of the
critical (closure) density, so that $\Omega = 1$ corresponds to a flat
universe as predicted by inflationary models \cite{11}. Finally, $h$ is the
present Hubble parameter in units of 100 km/(s$\cdot$Mpc); it lies in the
range $0.4 \leq h \leq 1$. The constraint that the universe be older than
10 billion years implies $\omeh \leq 1$; the true upper bound is almost
certainly tighter than this, but we want to be conservative since our
calculation will only be precise on the 10\% level.

In our calculation of the annihilation cross section $\sigma_{ij}$ of
eq.(\ref{e31}) we have only included annihilation into \ffbar\ pairs, where
$f$ is a fermion contained in the Standard Model. This under--estimates the
total annihilation cross section, and hence over--estimates the relic
density, if \lone\ is heavy enough to annihilate into pairs of gauge or Higgs
bosons, or mixed gauge--Higgs final states. However, in this case annihilation
into \ffbar\ final states is by itself usually sufficient to give an
acceptable relic density. The reason is that a large $|m_1|$ also implies
fairly large $SU(2)$ doublet components of \lone, and hence sizable couplings
to gauge and Higgs bosons.

The process $L_i^0 L_j^0 \to \ffbar$ can proceed via the exchange of a neutral
(scalar or pseudo--scalar) Higgs boson, or a $Z$ boson, in the $s-$channel. As
discussed in Sec.~3, there might also be terms in the superpotential that
couple exotic letpons to (s)fermions present in the MSSM. However, the
resulting contributions to the annihilation cross section are strongly
suppressed. We already saw that bounds on rare processes severely constrain
these couplings \cite{19}. Moreover, they would contribute to annihilation
only through the exchange of a sfermion in the $t-$ or $u-$channel. The
experimental lower bound on squark masses is already quite high \cite{27};
furthermore, as mentioned earlier, one expects sfermion masses to be of order
$M_{Z'}$ in this model \cite{5,6}, which leads to a strong suppression of
sfermion exchange contributions. Finally, these couplings involve the $SU(2)$
doublet components of $L_i^0$, which are usually quite small for $i=1,2$; the
corresponding contribution to the annihilation matrix element therefore
involves two small mixing factors. This is also true for the $Z-$exchange
contribution, but the $H L_i^0 L_j^0$ coupling (\ref{e21}) contains only one
small mixing angle if the Higgs boson is mostly an $SU(2)$
doublet.\footnote{Recall that this coupling results from the superpotential
(\ref{e1}), which couples one singlet to two doublets. An $SU(2)$ doublet
Higgs boson therefore couples to $L_i^0 L_j^0$ via one singlet and one doublet
component. Recall also that in this model all light Higgs bosons must be
dominantly doublets, since $x \gg v, \vb$ in eq.(\ref{e9}), so that the
singlet Higgs boson essentially decouples from physics at scale $M_Z$.} We
therefore neglect possible $t-$channel exchange diagrams.

Including the $s-$channel exchange of $Z$ and Higgs bosons, the relevant
matrix element can be written as
\be \label{e35}
{\cal A} \left( L_i^0 (k_1) L_j^0 (k_2) \to f(p_1) \bar{f}(p_2) \right) =
{\cal A}_{ij}^{(Z)} + {\cal A}_{ij}^{(A)} + {\cal A}_{ij}^{(h)},
\ee
with
\ben \label{e36} \beq
{\cal A}_{ij}^{(Z)} &= i \left( \frac {g} {\ctw} \right)^2 t_{ij} \bar{v}(k_2)
\gamma^\mu \gamma_5 u(k_1) \frac {g_{\mu\nu} - \frac {P_\mu P_\nu} {M_Z^2} }
{s - M_Z^2 + i M_Z \Gamma_Z} \bar{u}(p_1) \gamma^\nu ( a_f + b_f \gamma_5)
v(p_2); \label{e36a} \\
{\cal A}_{ij}^{(A)} &= i \frac {g} {\ctw} t_{ij}^{(A)} c^{(f)} \bar{v}(k_2)
\gamma_5 u(k_1) \frac {1} {s -m_A^2 + i m_A \Gamma_A} \bar{u}(p_1) \gamma_5
v(p_2); \label{e36b} \\
{\cal A}_{ij}^{(H)} &= i \frac {g} {\ctw} \bar{v}(k_2) u(k_1) \bar{u}(p_1)
v(p_2) \sum_{k=1}^3 \frac {t_{ijk}^{(h)} d_k^{(f)} } {s - m^2_{h_k^0} + i
m_{h_k^0} \Gamma_{h_k^0} }. \label{e36c}
\eeq \een
The $Z L_i^0 L_j^0$ couplings $t_{ij}$ are given in eq.(\ref{e7}), while the
$A L_i^0 L_j^0$ couplings $t_{ij}^{(A)}$ and the $h^0_k L_i^0 L_j^0$ couplings
$t_{ijk}^{(h)}$ can be read off eq.(\ref{e21}). The $A \ffbar$ couplings
$c^{(f)}$ and $h_k^0 \ffbar$ couplings $d_k^{(f)}$ are listed in
eqs.(\ref{e18a},\ref{e18b}). Finally, the $Z \ffbar$ couplings $a_f$ and $b_f$
are as usual given by $a_f = - \frac{1}{2} I_3^f + q^f \sin^2 \theta_W, \
b_f = \frac{1}{2} I_3^f$, with $I_3^f$ and $q^f$ the weak isospin and electric
charge of fermion $f$, respectively. In eq.(\ref{e36a}) we have introduced the
total momentum $P_\mu = (k_1+k_2)_\mu = (p_1+p_2)_\mu$, and $s = P^\mu P_\mu$.

The $A$ and $Z$ exchange diagrams interfere with each other, but not with the
scalar Higgs exchange contribution. The annihilation cross section can most
readily be computed using standard trace techniques. The result can be written
as
\be \label{e37}
\sigma \left( L_i^0 L_j^0 \to \ffbar \right) v = \frac {\beta_f N_f} {8 \pi s}
\left[ \wt{\sigma}_{ij}^{(ZZ)} + \wt{\sigma}_{ij}^{(ZA)} +
\wt{\sigma}_{ij}^{(AA)} + \wt{\sigma}_{ij}^{(hh)} \right],
\ee
where $\beta_f = \sqrt{1 - 4 m_f^2 /s}, \ N_f = 1 \ (3)$ for leptons (quarks),
and the scaled annihilation cross sections $\wt{\sigma}_{ij}$ are given by:
\ben \label{e38} \beq
\wt{\sigma}_{ij}^{(ZZ)} &= 4 \left( \frac {g}{\ctw} \right)^4 \left( t_{ij}
\right)^2 \left( a_f^2 + b_f^2 \right)
\left[ \frac { s^2/4 + s \beta_f^2 \vec{k}^2/3 - s (m_f^2 + m_i m_j)
+ 4 m_f^2 m_i m_j} { (s - M_Z^2)^2 + M_Z^2 \Gamma_Z^2}
\right. \nonumber \\ & \left. \hspace*{2.5cm}
+ \frac {2 m_f^2}{M_Z^2} \left( 1 - \frac {s} {2 M_Z^2} \right)
\frac { (m_i^2 - m_j^2)^2 } { (s - M_Z^2)^2 + M_Z^2 \Gamma_Z^2} +
\frac { m_f^2 (m_i+m_j)^2} {M_Z^4} \right]; \label{e38a} \\
\wt{\sigma}_{ij}^{(ZA)} &= 4 \left( \frac {g}{\ctw} \right)^2 t_{ij}
t_{ij}^{(A)} c^{(f)} b_f m_f (m_i+m_j) \frac {s-m_A^2}
{ (s - m_A^2)^2 + m_A^2 \Gamma_A^2} \frac {s-(m_i-m_j)^2} {M_Z^2};
\label{e38b} \\
\wt{\sigma}_{ij}^{(AA)} &= \left( \frac{g}{\ctw} \right)^2 \left( t_{ij}^{(A)}
c^{(f)} \right)^2 \frac {s - (m_i-m_j)^2} { (s - m_A^2)^2 + m_A^2 \Gamma_A^2};
\label{e38c} \\
\wt{\sigma}_{ij}^{(hh)} &= \left( \frac{g}{\ctw} \right)^2
\left[ s - (m_i+m_j)^2 \right] s \beta_f^2 \left| \sum_{k=1}^3 \frac
{t_{ijk}^{(h)} d_k^{(f)} } {s - m^2_{h_k^0} + i m_{h_k^0} \Gamma_{h_k^0} }
\right|^2 . \label{e38d}
\eeq \een
Here, the initial 3--momentum $\vec{k}^2$ is again given by eq.(\ref{e23})
with $M_Z^2$ replaced by $s$. As before, the masses $m_i$ and $m_j$ can have
either sign.

A few comments are in order. When writing the thermal average (\ref{e30}),
we have used non--relativistic kinematics; for consistency we therefore also
have to use a non--relativistic expression for $s$ in eqs.(\ref{e38}),
\be \label{e39}
s = \left( |m_i| + |m_j| \right)^2 + |m_i m_j | v^2.
\ee
this might seem dangerous, since in the presence of narrow poles the integral
in eq.(\ref{e30}) can receive sizable contributions from $v \sim 1$. However,
we have checked that using fully relativistic kinematics everywhere does not
change the result significantly; on the other hand, combining eq.(\ref{e30})
with a relativistic expression for $s$ can under--estimate the relic density
by a factor of 2 or 3.

In principle we now could compute \omeh\ numerically, by inserting
eqs.(\ref{e37}) and (\ref{e38}) into eqs.(\ref{e29})--(\ref{e34}). Note,
however, that inserting eq.(\ref{e30}) into eq.(\ref{e34}) leads to a double
integration. Since we want to test several million combinations of model
parameters, in order to make sure that we covered all relevant regions of
parameter space, a direct numerical integration is not practical. We used
the following approximate method instead.

As well known \cite{11,26}, the integral in eqs.(\ref{e30}) and (\ref{e34})
can be computed quite reliably from a simple analytical expression {\em if}
the annihilation cross section $\sigma_{\rm eff}$ does not depend too
sensitively on $v$. In this case one can use the Taylor expansion
\be \label{e40}
\sigma_{\rm eff} = a_{\rm eff} + b_{\rm eff} v^2,
\ee
which gives $\sigav (x) = a_{\rm eff} + 6 b_{\rm eff}/x$, and $J(x_f) =
a_{\rm eff} / x_f + 3 b_{\rm eff}/x_f^2$. Since $x_f \simeq 20$, annihilation
from an $s-$wave initial state, which contributes to $a_{\rm eff}$, reduces
the relic density more efficiently than annihilation from a $p-$wave does,
which only contributes to $b_{\rm eff}$. However, in our case this expansion
can be used with some reliability only for those contributions that do not
have an $s-$channel pole. Specifically, this includes the last term in the
squared $Z$ exchange contribution (\ref{e38a}), which is due to the
longitudinal polarization state of the $Z$; the $Z-A$ interference term
(\ref{e38b}); and the $H_k^0 - H_l^0$ interference terms in eq.(\ref{e38d}).
All other contributions do in general show strong variation with $v^2$, and
have to be treated separately.

Since we use the expansion (\ref{e40}) for all interference terms, we now only
need to compute the thermal average over a Breit--Wigner propagator,
multiplied with a power of $v^2$:
\be \label{e41}
I_n \equiv C \frac {x^{3/2}} {2 \sqrt{\pi}} \int_0^\infty dv e^{-x v^2/4}
\frac { (v^2)^{1+n} s(v) } { \left[ s(v) - m_P^2 \right]^2 + m_P^2
\Gamma_P^2},
\ee
where $s(v)$ is given by eq.(\ref{e39}), $C$ is a constant, and $P=Z, \ A$ or
$h_k^0$. Setting for simplicity $s= \left( |m_i| + |m_j| \right)^2$ in the
numerator (but not the denominator) of eq.(\ref{e41}), $I_n$ can be written as
\be \label{e42}
I_n = C \frac {x^{3/2}} {2 \sqrt{\pi}} \left( \frac {|m_i|+|m_j|} {m_i m_j}
\right)^2 \int_0^\infty dv e^{-x v^2/4} \frac {(v^2)^{1+n}} { \left( v^2 -
v_0^2 \right)^2 + \gamma},
\ee
with
\ben \label{e43} \beq
v_0^2 &= \frac {m_P^2 - \left( |m_i| + |m_j| \right)^2 } {|m_i m_j|};
\label{e43a} \\
\gamma &= \left( \frac {m_P \Gamma_P}{m_i m_j} \right)^2.
\label{e43b}
\eeq \een
Note that $v_0^2 < 0$ implies $s > m_P^2$ for all $v \geq 0$, so that the pole
is never accessible. On the other hand, if $v_0^2 > 0$, the contribution from
$v \simeq v_0$ to the integral in eq.(\ref{e42}) scales like $e^{-x v_0^2/4}
(v_0^2)^{1+n} \gamma^{-0.5}$, which can be substantial even for $v_0 \sim 1$
if $\gamma \ll 1$, i.e. if the pole is very narrow.

The integral in eq.(\ref{e42}) still seems to depend on three parameters (the
inverse temperature, and the position and width of the pole). Further progress
can be made by using the substitution $v' = \sqrt{x} v$:
\beq \label{e44}
I_n &= C \frac { x^{2-n}} {2 \sqrt{\pi}} \left( \frac {|m_i|+|m_j|} {m_i m_j}
\right)^2 \int_0^\infty dv' e^{-v'^2/4} \frac { (v'^2)^{1+n} }
{ \left( v'^2 - \tilde{v}_0^2 \right)^2 + \tilde{\gamma} }
\nonumber \\
&\equiv C x^{2-n} \left( \frac {|m_i|+|m_j|} {m_i m_j} \right)^2 \tilde{I}_n
(\tilde{v}_0^2, \tilde{\gamma}),
\eeq
with $\tilde{v}_0^2 = x v_0^2$ and $\tilde{\gamma} = x^2 \gamma$. We have
computed $\tilde{I}_0$ and $\tilde{I}_1$ numerically for 200 values of
$\tilde{v}_0^2$ between $-100$ and $+225$, and 50 values of $\tilde{\gamma}$
between $6.3 \cdot 10^{-3}$ and $3.5 \cdot 10^3$.\footnote{Recall that we need
to compute $I_n$ only for $x \geq 20$. Moreover, $\Gamma_P/m_P \geq 10^{-4}$
even for the light Higgs bosons of the model.} Note that these 10,000 values
of $\tilde{I}_0$ and $\tilde{I}_1$ need to be computed only once; afterwards
the thermal average over the annihilation cross section can be computed
without numerical integration: If $\tilde{v}_0^2$ does not lie in this range,
the pole is so distant that the expansion (\ref{e40}) can be used for it; for
values of $\tilde{v}_0^2$ and $\tilde{\gamma}$ in the specified range, the
$\tilde{I}_n$ are estimated by interpolation.

Specifically, for $i=j \ (\lone\lone$ or \ltwo\ltwo\ annihilation), only the
squared $A$ exchange term (\ref{e38c}) gets contributions $\propto
\tilde{I}_0$ when inserted in eq.(\ref{e30}); the resonant contributions to
eq.(\ref{e38a}), as well as eq.(\ref{e38d}), are proportional to $v^2$ for
$m_i = m_j$, i.e. only contribute via $\tilde{I}_1$. The reason is that
on--shell $Z$ and scalar Higgs bosons can only be produced from two identical
Majorana fermions if they are in a $p-$wave state, while on--shell
pseudo--scalar bosons can be produced from an $s-$wave state. However, squared
$Z$ and scalar Higgs exchange do contribute to $\tilde{I}_0$ terms if $m_i
\neq m_j$. In the important special case where \lone\ and \ltwo\ form a Dirac
fermion, one has $m_1 = -m_2$; in this case squared $A$ exchange only starts
at order $v^2$ ($\tilde{I}_1$ terms only), while squared scalar Higgs and $Z$
exchange start at order $v^0$. Finally, for all contributions that start at
order $v^0$ ($\tilde{I}_0$ terms), we have expanded the numerator to order
$v^2$, i.e. added a (properly normalized) $\tilde{I}_1$ term. We checked
numerically that our combination of eq.(\ref{e40}) for non--resonant
(interference) terms, and eq.(\ref{e44}) with interpolation to the actual
values of $\tilde{v}_0^2$ and $\tilde{\gamma}$, reproduces the exact thermal
average of the annihilation cross section to an accuracy of about 10\% or
better; this is quite sufficient for us. In contrast, simply using the
expansion (\ref{e40}) for the entire cross section can both over-- and
under--estimate the true thermal average by a large factor; this had been
observed previously \cite{28} in the similar case of the MSSM with scalar
Higgs mass close to twice the LSP mass. Finally, the annihilation integral
(\ref{e34}) usually converges rather quickly; we have therefore computed it
numerically, still using the method outlined above to determine $\sigav(x)$ in
the integrand.

\setcounter{footnote}{0}
\section*{5) Results}
We are now in a position to describe our numerical studies of the model, using
the expressions given in the previous three sections. Our basic approach is to
randomly sample the parameter space, and to count the number of solutions that
satisfy all the constraints we impose, including those that can be derived
from searches at future (linear) \epem\ colliders. The ultimate goal is to
devise a set of constraints such that there are no acceptable solutions left.
This means that either a signal characteristic for the model has been found,
i.e. one of the hypothetical future searches is successful, or the model is
completely excluded, independent of the values of the free parameters.

We chose this Monte Carlo approach since the number of free parameters is too
large to allow for a systematic scan of the entire parameter space. The mass
matrix (\ref{e3}) for the neutral exotic leptons already contains ten free
parameters. For each scan of parameter space, we impose a lower bound on the
charged leptons masses $m_{L_i^\pm}$, since these particles should be trivial
to discover at an \epem\ collider unless their production is kinematically
suppressed. We also fix the ratio $\tanb \equiv v/\vb$ for a given run; this
determines $v$ and \vb, since $\sqrt{v^2+\vb^2} = 2 M_W/g$ is known. The
Yukawa couplings appearing in eq.(\ref{e3}) are then chosen randomly in the
interval $|\la_{ijk}|\leq \lmax$. We take $\lmax=0.85$, which is approximately
the upper bound on any one coupling from the requirement that there should be
no Landau pole at scales below the GUT scale. This is a conservative approach,
since this bound is significantly stronger if several Yukawa couplings are
sizable \cite{10}, which is required if \ml\ is to be close to its upper
bound.

Having specified the exotic lepton sector, we check whether the present LEP
constraints (\ref{e24}) and (\ref{e25}) are satisfied. If so, we check whether
the total ``visible" exotic cross--section, determined from eq.(\ref{e26}),
is sufficiently small so that a given \epem\ collider will not detect pair
production of exotic leptons; this constraint obviously depends on the
collider we are considering, as described below. Recall that we consider a
neutral exotic lepton to be ``visible" only if it is at least 5 GeV heavier
than \lone.

This exhausts the constraints from searches for exotic leptons at present and
future \epem\ colliders. If our choice of parameters is still viable, we next
randomly pick a Higgs sector, subject to the constraints (\ref{e27}). Since
present bounds on the mass of the $Z'$ boson are already quite high \cite{2},
we always find that there is one singlet--like neutral Higgs boson, which
plays no role in any of the process we are considering; it is therefore
sufficient to simply fix the SM singlet vev $x$ to some large value, say 2
TeV. Since our runs are for fixed \tanb, we only need to chose the values of
two additional parameters in order to completely specify the Higgs sector of
the model, see eq.(\ref{e8}). We chose these to be the coupling $\la_{333}$,
which also has to lie in the interval $|\la_{333}| \leq 0.85$, and the mass
$m_A$ of the physical pseudo--scalar Higgs boson; this then fixes the
parameter $A_{333}$ via eq.(\ref{e12}). Note that for each set of leptonic
parameters that satisfy constraints from \epem\ colliders, we keep chosing
pairs $(\la_{333}, m_A)$ at random until we have found an acceptable Higgs
sector.

This is necessary because Higgs exchange contributions can play an important
role in the calculation of the \lone\ relic density. Once the exotic lepton
sector and the Higgs sector are specified, all quantities appearing in the
annihilation cross sections (\ref{e38}) are fixed, and we can compute the
relic density as described in Sec.~4, and check whether it is acceptable.

Figs.~1a,b show that present constraints [the bounds (\ref{e24}), (\ref{e25})
and (\ref{e27}), together with $\omeh \leq 1$] are still quite far from
testing the model decisively. For these figures we have allowed charged lepton
masses as low as 40 GeV when sampling the parameter space; we have however
required $|m_{L_i^\pm}| \geq 45$ GeV for all acceptable solutions. This
explains why the survival fraction after imposing LEP1 constraints only is
less than unity in some bins with $\ml > M_Z/2$.

Note that LEP1 searches only impose weak constraints for very small values of
\ml. The reason is that most parameter sets with small \ml\ have large masses
for the charged exotic leptons. In this case four of the six neutral exotics
are also heavy, while the remaining two states are mostly SM singlets, i.e.
couple only weakly to the $Z$ boson. However, the relic density constraint is
most effective precisely in this situation, since it leads to small
annihilation cross sections: The couplings of singlet--like exotics to the
light Higgs bosons are also weak, and their small masses suppress the cross
sections even further; far below the pole, $s-$channel exchange contributions
scale like $m^2_{L_i^0}/m_P^4$, where $P$ is the exchanged particle. We
therefore find no cosmologically acceptable scenario with $\ml \leq m_b \simeq
5$ GeV; this is not surprising since the $b \bar b$ final state, if
accessible, contributes most to Higgs exchange diagrams.

Figs.~1 also reveal a technical problem: even though each figure is based on
$10^5$ sets of leptonic parameters (``hundred thousand models"), the region of
large \ml\ is only sparsely populated. This is perhaps not surprising, since
all ten parameters appearing in the mass matrix (\ref{e3}) must be chosen
within a narrow range if \ml\ is to come out close to its upper bound. At the
same time, combinations of parameters leading to large \ml\ are most difficult
to exclude, i.e. most easily evade all bounds. For one thing, large \ml\
implies large $SU(2)$ doublet components of \lone, and hence large
annihilation cross sections and a small relic density. Further, constraints
from collider searches can most easily be evaded if \ltwo\ is close in mass to
\lone, since \ltwo\ then also becomes effectively invisible. In such a
situation $\lone \ltwo$ co--annihilation can also reduce the relic density
even further. Since our goal is to test the model decisively, it would be
advantageous if in our sampling of parameter space we could give preference to
regions that are most difficult to exclude.

Fortunately a large \ml\ is correlated with a small $\lone-\ltwo$ mass
difference. In ref.\cite{10} it has been shown that choices of parameters that
maximize \ml\ always lead to a situation where the six eigenvalues of the mass
matrix (\ref{e3}) come in three pairs that only differ by a sign; in such a
situation the six Majorana states can also be described by three neutral Dirac
fermions. In particular, the Majorana state \ltwo\ will now be completely
invisible, being degenerate in mass with \lone\ and hence (by assumption)
stable; this obviously also maximizes $\lone \ltwo$ co--annihilation. From now
on we will therefore only show results for the subset of parameter space where
the six neutral exotic leptons do indeed form three Dirac fermions. This can
be enforced by chosing
\be \label{e45}
\la_{131} = \la_{132}, \ \la_{231} = \la_{232}, \ \la_{311} = - \la_{312},
\ \la_{321} = - \la_{322}.
\ee
For given \tanb, this reduces the number of parameters in the exotic lepton
sector from ten to six; this reduced parameter space is obviously much easier
to sample exhaustively. We have checked that our runs with eqs.(\ref{e45})
imposed do always find significantly more acceptable solutions than scans of
the entire parameter space with equal statistics.

Figs.~2a,b show that, at least if the mass matrix (\ref{e3}) has Dirac
structure, neutral lepton searches at LEP2 will not lead to significant new
constraints, unless a light charged exotic is found. We have assumed here that
LEP2 will reach a center--of--mass energy \rs\ of 190 GeV, and that the
production of neutral leptons is detectable if the cross section, summed over
all ``visible" modes, exceeds 20 fb, which corresponds to 10 events per
experiment for the foreseen integrated luminosity of 500 pb$^{-1}$. Comparison
with Figs.~1 shows that the fraction of parameter space excluded by present
LEP1 constraints has become smaller, even for light \lone. This is mostly
because $\lone \ltwo$ and $\ltwo \ltwo$ final states are now invisible. Note
also that we now find some cosmologically acceptable solutions with
$\tanb=1.2$ and \ml\ as small as 10 GeV; this indicates that co--annihilation
can indeed reduce the relic density significantly. Further, even though the
absolute upper bound on \ml\ decreases with increasing mass of the charged
exotic leptons \cite{10}, Figs.~2a,b extend to larger values of \ml\ than
Figs~1a,b do; clearly the restrictions (\ref{e45}) have made it much more
likely to produce scenarios with large \ml. Finally, comparison of Figs. 2a
and 2b shows that LEP constraints exclude a larger fraction of parameter space
for large \tanb; this can already be seen from Figs.~1a,b. We showed in Sec.~2
that increasing \tanb\ decreases the upper bound on \ml, since $\det {\cal
M}_{L^0} \propto (v \vb x)^2$; however, the size of the $SU(2)$ doublet
components of the light exotics, and hence their couplings to the $Z$ boson,
remains more or less the same. Reducing the masses of the exotics while
leaving their couplings essentially unchanged obviously increases the partial
width for $Z$ decays into exotic leptons.

Figs.~2 clearly show that LEP2 will not be able to test the model decisively;
there are many choices of parameters that lead to an acceptable cosmology, but
no ``new physics" signal at this collider. One will therefore need (linear)
colliders operating at higher energies in order to probe the entire parameter
space. Such colliders are often assumed to be built in three stages, where the
energy is increased from about 0.5 TeV to 1.0 TeV and, eventually, 1.5 TeV or
even higher. We generically call these three stages NLC1, NLC2 and NLC3.

Figs.~3a,b show the situation if a 500 GeV \epem\ collider fails to discover
pair production of exotic leptons, which we have interpreted as meaning that
the total cross section into visible exotic final states is less than 0.5 fb;
the integrated luminosity of such a collider is usually assumed to be several
tens of fb$^{-1}$ per year. For these runs we have required the masses of the
charged exotic leptons to exceed 240 GeV. This reduces both the upper bound on
\ml\ and the maximal $SU(2)$ doublet component of \lone\ significantly. As a
result, for $\tanb=1.2$ present LEP1 searches do not constrain the model any
further. For $\tanb=5$, these constraints still do exclude some combinations
of parameters, but they are clearly much less restrictive here than for lower
masses of the charged exotics, see Fig.~2b. On the other hand, unlike at LEP2,
searches for neutral exotics at NLC1 can probe sizable regions of parameter
space even if no charged exotic leptons are found; this can be seen from the
large differences between the dotted and dashed histograms in Figs.~3.
Finally, the relic density constraint again excludes parameter choices giving
$\ml \leq m_b$, and impose significant constraints as long as $\ml \leq 15$ to
20 GeV. However, a substantial number of parameter sets still satisfies all
constraints (solid histograms).

This remains true even if a 1 TeV collider fails to discover pair production
of exotic leptons. Indeed, Figs.~4a,b show that a small region of parameter
space survives even if searches for pairs of exotic leptons at a 1.5 TeV
collider remain unsucessful; here we have assumed that the integrated
luminosity scales like the square of the beam energy, so that a signal
exceeding 0.05 fb would be detectable. We evidently need to find some
additional constraint(s) if we want to test the model decisively.

There are significant differences between Figs.~4a,b and 3a,b, which might
offer a clue as to what these additional constraints might be. To begin with,
the minimal allowed value of \ml\ for small \tanb\ has increased from about
6 GeV (Fig.~3a) to about 23 GeV (Fig.~4a). Unsuccessful earches at a 1.5 TeV
collider force the $SU(2)$ doublet components of \lone\ to be much smaller
than searches at an 0.5 TeV collider do, which leads to considerably reduced
couplings of the light exotic leptons to gauge and Higgs bosons. This has to
be compensated by an increase of \ml\ in order to keep the relic density
acceptably small; recall that for light \lone, the annihilation cross sections
scale like $m^2_{L_1^0}$. Even more importantly, if a 1.5 TeV collider fails
to find evidence for the pair production of exotic leptons, the upper bound
on \ml\ is reduced to about 30 (11) GeV for $\tanb=1.2$ (5), as compared to
84 (27) GeV after unsuccessful searches at NLC1.

Recall that the masses of the two lightest exotic Majorana states, or the
lightest exotic Dirac state if eqs.(\ref{e45}) hold, come from vevs that break
$SU(2)$. This indicates that there is a correlation between \ml\ and the size
of the couplings between light exotics and $SU(2)$ doublet Higgs bosons,
although the relation is not as simple as that between the masses and Yukawa
couplings of the quarks and leptons of the SM. Further, Higgs searches at LEP1
imply that the mass of a scalar Higgs boson with unsuppressed $ZZH$ coupling
must exceed 60 GeV, see eq.(\ref{e27a}). Such a Higgs boson can therefore
decay into pairs of light neutral exotic leptons for all allowed combinations
of parameters shown in Figs.~4; this could lead to a large invisible branching
ratio of this Higgs boson. Indeed, we find that all surviving scenarios shown
in Fig.~4a have one rather light neutral scalar Higgs boson, with essentially
unsuppressed coupling to two $Z$ bosons, and with invisible branching ratio
exceeding 80\%. Moreover, it will be produced copiously at any \epem\ collider
with $\rs \geq 300$ GeV, since its mass cannot exceed 150 GeV or so, see
eq.(\ref{e10}). Note that such a Higgs boson is easily detectable even if it
decays invisibly \cite{29}, since it would be produced in association with a
$Z$ boson, whose decay products would be sufficient to reconstruct $m_H$.
Finally, a large invisible branching ratio for a light scalar Higgs boson
cannot be accommodated in the SM; in the MSSM it would imply the existence of
a chargino light enough to be discovered at LEP2 \cite{24}. It therefore
constitutes a signal for the kind of model we are considering.

The situation at large \tanb, Fig.~4b, is somewhat more complicated. Notice
that, unlike in Fig.~4a, now only a small fraction of all parameter sets that
give \ml\ close to its upper bound, and do not lead to a detectable signal for
exotic lepton pair production, survives the relic density constraint. The
reason is that for small \tanb\ and with \ml\ close to its upper bound, $Z$
exchange by itself is usually sufficient to give an acceptable relic density.
On the other hand, increasing \tanb\ to 5 reduces the maximal \ml\ so much
that $Z$ exchange, the cross section for which scales like
$m^2_{L_1^0}/M_Z^4$, is no longer sufficient; light Higgs bosons have to be
present to enhance the annihilation cross section. In particular, a light
pseudo--scalar Higgs boson $A$ allows $\lone \lone$ and $\ltwo \ltwo$ to
annihilate from an $s-$wave initial state, which gives a larger thermal
average than annihilation from a $p-$wave initial state does, as discussed
below eq.(\ref{e40}) in Sec.~4.\footnote{\lone\ltwo\ co--annihilation via
scalar Higgs exchange could come from an $s-$wave initial state; however, if
the neutral lepton mass matrix has Dirac structure, the off--diagonal $\lone
\ltwo h_i^0$ couplings vanish identically, as does the $\lone \ltwo A$
coupling.} Therefore even if \ml\ is close to its upper bound, the relic
density will only be acceptable if $m_A$ is chosen to be fairly small, which
is true only for some fraction of Higgs parameter space. We find that all
surviving parameter sets in Fig.~4b have $m_A < 110$ GeV, and also have two
neutral scalar Higgs bosons with masses below 125 GeV. At least one of these
three light neutral Higgs bosons has invisible branching ratio exceeding 50\%.
All three of these Higgs bosons will be produced copiously, either in
association with a $Z$ boson or as $A h_i^0$ pairs. A large invisible
branching ratio for any of these Higgs particles is therefore again a
distinctive signature. Note that now the invisible branching ratio is always
less than 75\%, so that in at least 50\% of all $A h_i^0$ pairs at least one
of the two Higgs bosons will decay into a visible final state, mostly $b \bar
b$ pairs. This not only guarantees the detectability of this final state, but
also ensures that the invisible branching ratios can be measured with some
accuracy, e.g. by comparing the rate for events with one invisible Higgs boson
(single--sided events) with that for events where both Higgs particles leave a
detectable final state. Finally, we note that in the surviving cases the
charged Higgs boson mass (\ref{ech}) also is below 125 GeV, and usually below
100 GeV; such a light charged Higgs boson is already almost excluded in the
MSSM, and might thus give a second indication for physics beyond the MSSM.

Based on the results of Figs.~4 we therefore conclude that in this model
either a signal for the pair production of exotic leptons will be observed, at
the latest at an \epem\ collider with $\rs=1.5$ TeV; or at least one neutral
Higgs boson with mass below 150 GeV and invisible branching ratio exceeding
50\% will be found at any \epem\ collider with $\rs \geq 300$ GeV. If neither
of these two signals is detected, the model can be completely excluded.

Since this conclusion is based on a large but finite Monte Carlo sampling of
parameter space (Figs.~4a,b contain $5 \cdot 10^5$ sets of leptonic parameters
each), we have checked that it also holds in a ``worst case" scenario. To this
end we have chosen the parameters entering the mass matrix (\ref{e3}) such
that \ml\ is maximized \cite{10}:
\ben \label{e46} \beq
\la_{131} &= \la_{132} = \la_{311} = - \la_{312} = \la_{231} = - \la_{321}
= \la_{232} = \la_{322} = \lmax; \label{e46a} \\
m_{L_1^\pm} &= - m_{L_2^\pm}. \label{e46b}
\eeq \een
Note that this ansatz is consistent with eqs.(\ref{e45}). In Fig.~5 we show
the resulting maximal value of \ml\ as a function of \tanb, subject to the
constraints $m_{L_1^\pm} > m_{L^\pm,{\rm min}}$ and $|\ml| + |m_{L_3^0}| > 2
m_{L^\pm,{\rm min}}$. The latter bound approximates the constraint that can be
derived from an unsuccessful search for associate $\lone L_3^0$ production;
note that the relevant coupling is always quite large for the ansatz
(\ref{e46}).\footnote{Recall that the production of two light neutral exotic
leptons leads to an invisible final state.} For our choice $\lmax=0.85$, the
maximal allowed \ml\ for $m_{L^\pm,min} = 700$ GeV is indeed around 30 GeV, in
agreement with results shown in Fig.~4a; for $\tanb=5$, $m_{L_1^0,max}$ falls
to about 11 GeV, in agreement with Fig.~4b. We should mention here that the
choice $\lmax = 0.85$ in eq.(\ref{e46a}) is very conservative. In
ref.\cite{10} it has been pointed out that $\lmax > 0.7$ implies the existence
of a Landau pole at a scale below $10^{10}$ GeV, the smallest energy scale
where one might expect the gauge group (and hence the relevant renormalization
group equations) to change in this class of models. Note that for large
$m_{L^\pm,{\rm min}}$, the upper bound on \ml\ scales like the square of
\lmax; reducing \lmax\ to 0.7 therefore means that $\ml \leq 30$ GeV already
for $m_{L^\pm,{\rm min}} = 480$ GeV, in which case a 1 TeV collider would be
sufficient to test the model decisively in the manner described
above.\footnote{The larger value of \lmax\ we are using here also explains why
Fig.~5 allows larger values of \ml\ than ref.\cite{10} does.}

We also checked whether it is possible to chose parameters such that the
SM--like Higgs boson only has very weak couplings to light neutral exotics.
This is indeed possible, but only for sizable \tanb, and only if the
pseudoscalar Higgs boson is quite light. In this case the model contains two
light $SU(2)$ doublet neutral scalar Higgs bosons, as mentioned earlier; the
``SM--like Higgs" is defined to be the one with the larger coupling to two $Z$
bosons. We saw in Fig.~4b that for large \tanb, the relic density constraint
can only be satisfied if $m_A$ is rather small; the conditions for a scenario
with small invisible width of the SM--like Higgs are therefore satisfied.
However, Fig.~6 shows that either the pseudoscalar Higgs or the lightest
neutral scalar can always decay into an \lone\lone\ pair if searches at a 1.5
TeV collider do not find a signal for the pair production of exotic leptons.
In this figure we show the minimal allowed values of $m_A$ and of $m_{h_1^0}$,
as well as the minimum of the sum of these two masses, as a function of \tanb,
where we have only used the present LEP1 constraints (\ref{e27}). For $\tanb
\simeq 1.5$, $m_A$ could be as low as 34 GeV, so that $A \to \lone\lone$
decays would be kinematically forbidden over a sizable region of parameter
space. However, in this case the lightest scalar Higgs boson must be heavier
than 50 GeV, since $m_A + m_{h_1^0} > 84$ GeV. Note that in this scenario,
$h_1^0$ is usually not the SM--like Higgs, and does have sizable couplings to
light exotic leptons. Moreover, we find that the invisible branching ratio of
the SM--like scalar Higgs can only be reduced to a value below 10\% if $\tanb
\geq 3$, in which case unsuccessful searches for lepton pair production at
NLC3 imply $\ml < 20$ GeV, see Fig.~5. In this case both $A$ and $h_1^0$ have
large invisible branching ratios; remember that a sizable contribution from
$A$ exchange to \lone\ annihilation, and hence a substantial $A \lone \lone$
coupling, is required to satisfy the relic density constraint for $\tanb > 2$.
Finally, we found no acceptable solutions where the invisible branching ratio
of a light Higgs boson can be diluted by decays into pairs of even lighter
Higgs bosons ($h_2^0 \to h_1^0 h_1^0$ or $h_2^0 \to AA$). We conclude that it
is indeed impossible to devise a Higgs sector such that no light Higgs boson
has large invisible branching ratio if NLC3 fails to find a signal for the
pair production of exotic leptons.

Finally, we investigated the question whether Higgs searches at LEP2 will
allow us to sharpen our predictions. This does not seem to be the case, at
least as far as searches at a 1.5 TeV collider are concerned. As mentioned
earlier, for small \tanb, $Z$ exchange by itself can be sufficient to give an
acceptable relic density, so increasing the experimental lower bounds on Higgs
masses will have little effect on Fig.~4a. For $\tanb \geq 3$ one needs $m_A
\leq 110$ GeV in order to achieve $\omeh \leq 1$. This upper bound is almost
independent of \tanb\ once it exceeds 3 or so; the decrease of \ml\ caused by
increasing \tanb\ is balanced almost perfectly by the increase of the $A b
\bar b$ and $A \tau^+ \tau^-$ couplings, leading to a constant $A$ exchange
contribution, as long as $\ml > m_b$. However, this bound still allows values
of $m_A$ just beyond the reach of LEP2.

On the other hand, if LEP2 can increase the lower bound on the SM Higgs to 90
GeV or more, a 1 TeV \epem\ collider would be sufficient to test the model
decisively even with our choice $\lmax = 0.85$, since the absence of a signal
for exotic lepton pair production at such a collider implies $\ml < 45$ GeV,
see Fig.~5; one would then again be in the situation where at least one
copiously produced Higgs boson has to have a large invisible branching ratio.
Unfortunately in the absence of a positive signal for lepton pair production,
a 0.5 TeV collider would never be sufficient to test the model uniquely,
since, as shown in Fig.~3a, if would only allow to establish the bound $\ml <
85$ GeV, well above half the maximal allowed mass of the SM--like Higgs boson.
One can then always find cosmologically acceptable scenarios where only one
Higgs boson is experimentally accessible, with small or vanishing invisible
branching ratio; such a model would be indistinguishable from the MSSM, or
even the SM, as far as the NLC1 is concerned.

\section*{6) Summary and Conclusions}

In this paper we have shown that searches for exotic leptons and Higgs bosons
at future \epem\ colliders, when combined with cosmological constraints, can
test a large class of $E(6)$ models decisively. The basic idea is that, due to
the structure of the mass matrix for the 6 neutral exotic leptons predicted by
the model, the failure to detect the production of heavy (charged or neutral)
exotic leptons at such colliders will allow to derive increasingly stronger
upper bounds on the masses of the light neutral exotics. Eventually these
leptons will have to be lighter than half the mass of the light neutral Higgs
bosons of the model. The only way to suppress the couplings of these light
exotic leptons to all light Higgs bosons is to make the leptons almost pure
$SU(2)$ singlets. However, in this case exotic leptons produced in the very
early universe would have small annihilation cross sections, so that their
present relic density would be unacceptably large. All cosmologically
acceptable parameter sets therefore predict that either the pair production of
exotic leptons is visible at a 1.5 TeV \epem\ collider, or at least one light
neutral Higgs boson will have an invisible branching ratio exceeding 50\%; if
LEP2 fails to discover a Higgs boson, a 1 TeV \epem\ collider would be
sufficient.

In order to arrive at this conclusion, we had to make two crucial assumptions.
First, no $SO(10)$ singlet scalar $N$ is allowed to have a vev at an
``intermediate scale" of ${\cal O}(10^{10})$ GeV or more. This assumption is
not unreasonable, since such a large vev would allow to push the masses of
almost all new particles predicted by $E(6)$ up to this high scale, including
the new gauge bosons, the exotic quarks and leptons, and their superpartners.
Apart from the possible presence of light right--handed (s)neutrinos, the
model would then look like the MSSM at scales below this vev. Note that we do
allow $SO(10)$ nonsinglets to have such a large vev. Secondly, we have to
assume that $R-$parity is conserved. Otherwise the lightest exotic lepton,
which has odd $R-$parity, could decay even if it is lighter than the lightest
neutralino, in which case the relic density constraint would be satisfied
trivially. However, if $R-$parity is broken, this rather complicated model
would not be able to accommodate cold dark matter; recent attempts to
understand structure formation in the universe strongly favor scenarios with a
substantial amount of cold dark matter \cite{25a}.

We also made a few additional assumptions in order to simplify our
calculation. However, since we have been very conservative in our
interpretation  of the cosmological constraint, and of the upper bound on the
Yukawa couplings of the exotic leptons that follows from the requirement that
no coupling should have a Landau pole below the intermediate scale of
$10^{10}$ GeV, we believe that our final result holds even if these
additional, technical assumptions are relaxed. Specifically, we have assumed
that the exotic leptons do not mix with charginos and neutralinos. This
assumption is technically natural. Moreover, in ref.\cite{10} it has been
argued that, while such mixing might increase the upper bound on \ml\ for {\em
fixed} Yukawa couplings, it would also necessitate the existence of additional
terms in the superpotential; these would lower the upper bound on the relevant
Yukawa couplings from the absence of Landau poles. The total change of the
upper bound on \ml\ is therefore quite modest.

Further, when computing the annihilation cross sections, we have assumed that
there are no terms in the superpotential which couple exotic leptons to
ordinary quark and lepton superfields. However, we argued in Sec.~4 that such
couplings would in any case contribute negligibly to the total annihilation
cross section. Finally, we have taken the $Z'$ boson(s) of the model to be too
heavy to contribute either to the pair production or to the annihilation of
exotic leptons. Present lower bounds on $M_{Z'}$ already ensure that $Z'$
exchange contributions to the annihilation cross sections are negligible. They
could still affect the production of exotic leptons at TeV--scale \epem\
colliders, but in this case the existence of a $Z'$ boson could also be
inferred from studies of ordinary quark and lepton pair production, which
would again give a good signal for the class of models we are studying.
Besides, the constraints on the exotic lepton sector that could be derived
from new particle searches at future colliders are likely to be more, not
less, severe if there is a significant contribution from $Z'$ exchange, since
even the $SU(2)$ singlet fields $\wt{N}$ couple to the $Z'$ boson with full
gauge strength.

There is one more assumption that we have not mentioned so far: when using the
formalism described in Sec.~4 to estimate the relic density, we have assumed
that the exotic leptons are non--relativistic (``cold") when they drop out of
thermal equilibrium. This is true if their masses exceed a few hundred MeV or
so, and the relic density constraint will remain valid down to much smaller
masses, in the keV range, but our calculation clearly breaks down if the light
exotics are (nearly) massless. This could, for example, be achieved by setting
all couplings in the superpotential (\ref{e1}) to zero, except for $\la_{113}$
and $\la_{223}$ which are needed to give masses to the charged exotic leptons.
The two lightest neutral exotics would then be massless $SU(2) \times U(1)_Y$
singlets. However, this would increase the density of relativistic particles,
and hence the expansion rate of the universe, in the epoch when light nuclei
are formed. The most recent analysis \cite{30} finds that data seem to favor
models where the effective number of SM neutrinos is smaller, not larger,
than three. A model with additional massless fermions is therefore strongly
disfavored.

Finally, it can be argued that the upper bound (\ref{e10}) on the mass of the
lightest neutral scalar Higgs boson of the model allows a much easier test.
However, a very similar bound also holds in the MSSM and, indeed, in all SUSY
models where the Higgs sector is required to remain perturbative up to some
high scale \cite{31,18}. This test would therefore not be very specific. While
the failure to detect such a Higgs boson would rule out a very large class of
models, including the $E(6)$ models we are considering here, discovery of the
Higgs boson may not be sufficient to distinguish between the present model and
the MSSM, or even the non--supersymmetric SM. In contrast, the large mass
splitting between the light and heavy exotic leptons implies that exotic
lepton production should not be confused with the production of charginos and
neutralinos predicted by the MSSM. Large invisible branching ratios for light
Higgs bosons also clearly indicate the presence of (super)fields beyond those
contained in the (MS)SM, unless the mass of the light chargino is close to its
present lower bound. Singlet Majoron models \cite{32} can also have light
Higgs bosons with large invisible branching ratios \cite{33}. However, in such
models the light bosons have sizable $SU(2)$ singlet components, in contrast
to the models we are studying here, where the large mass of the $Z'$ boson
forces all light Higgs bosons to be predominantly $SU(2)$ doublets. A detailed
study of the production and decay of the light Higgs boson(s) should therefore
be able to distinguish between the $E(6)$ model and the singlet Majoron model.

Finally, we would like to emphasize that the decisive test we have devised
here relies on the versatility of high energy \epem\ colliders. Clearly the
cross sections we are studying are too small to give viable signals at hadron
colliders. Moreover, the search for the production of neutral leptons, in
particular the associate production of a light and a heavy neutral exotic,
plays an important role in our analysis. This would not be feasible at $e
\gamma$ or $\gamma \gamma$ colliders, which may also be unable to discover,
let alone study, Higgs bosons with large invisible branching ratios. All these
colliders will be able to impose constraints on, or -- with luck -- to
discover, some of the new particles predicted by $E(6)$ models. However, only
searches at \epem\ colliders, when combined with cosmological considerations,
seem capable of excluding these models completely.

\subsection*{Acknowledgements}
The work of M.D. was supported in part by the U.S. Department of Energy under
grant No. DE-FG02-95ER40896, by the Wisconsin Research Committee with funds
granted by the Wisconsin Alumni Research Foundation, as well as by a grant
from the Deutsche Forschungsgemeinschaft under the Heisenberg program.


\newpage
\section*{Figure Captions}
\renewcommand{\labelenumi}{Fig. \arabic{enumi}}
\begin{enumerate}

\vspace{4mm}
\item
The fraction of parameter sets that survive certain constraints is shown as a
function of the mass \ml\ of the lightest neutral exotic lepton, for
$\tanb=1.2$ (a) and 5 (b). The fraction is relative to the total number of
sets generated in a given bin. For the dotted histograms only the current,
LEP1 constraints (\ref{e24}),(\ref{e25}) have been imposed, while the solid
histograms show the fraction of parameter sets that in addition satisfy $\omeh
< 1$. Recall that we only include $f \bar f$ final states in the calculation
of the relic density; this under--estimates the annihilation cross sections,
i.e. over--estimates the relic density, for $\ml > m_W$. When chosing the
parameters of the neutral lepton mass matrix (\ref{e3}), we have required the
charged exotic leptons to be heavier than 40 GeV.

\vspace{4mm}
\item
The fraction of parameter sets that survive present LEP1 constraints (dotted),
exotic lepton searches at LEP2 (dashed), and the relic density bound $\omeh <
1$ (solid). We have assumed that LEP2 will be able to detect pair production
of neutral exotic leptons if the total cross section for final states where at
least one leptons is more than 5 GeV heavier than \lone\ exceeds 20 fb. For
reasons explained in the text, we have imposed the restrictions (\ref{e45}) on
the parameters of the neutral lepton mass matrix (\ref{e3}), which implies
that the 6 neutral Majorana leptons pair up to form 3 Dirac states. We have
required the charged exotic leptons to be heavier than 80 GeV here. Results
are for $\tanb=1.2$ (a) and 5 (b).

\vspace{4mm}
\item
As in Fig.~2, except that we have required the charged exotic leptons to be
heavier than 240 GeV, and the dashed histograms show the fraction of parameter
sets that would not lead to an observable signal for the pair production of
exotic leptons at a 500 GeV \epem\ collider (``NLC1"). Here a signal is
considered to be observable if the total cross section for the production of
neutral exotic leptons, with one final state lepton being at least 5 GeV
heavier than \lone, exceeds 0.5 fb.

\vspace{4mm}
\item
As in Fig.~2, except that we have required the charged exotic leptons to be
heavier than 700 GeV, and the dashed histograms show the fraction of parameter
sets that would not lead to an observable signal for the pair production of
exotic leptons at a 1.5 TeV \epem\ collider (``NLC3"). Here a signal is
considered to be observable if the total cross section for the production of
neutral exotic leptons, with one final state lepton being at least 5 GeV
heavier than \lone, exceeds 0.05 fb. As explained in the text, all surviving
parameter sets (solid histograms) predict a large invisible branchig ratio for
at least one light neutral Higgs boson, which allows to test the model
decisively.

\vspace{4mm}
\item
The maximal value of the mass \ml\ of the lightest neutral exotic lepton,
computed from the ansatx (\ref{e46}) with $\lmax=0.85$, is shown as a function
of \tanb. We have required $m_{L_i^\pm} > m_{L^+,{\rm min}}$, and $|\ml| +
|m_{L_3^0}| > 2  m_{L^+,{\rm min}}$, for different values of $ m_{L^+,{\rm
min}}$ as indicated.

\vspace{4mm}
\item
The minimal masses of the Higgs bosons of the model, as determined from
present LEP1 constraints (\ref{e27}), is shown as a function of \tanb. The
solid, dashed and dotted curves show the smallest allowed mass of the lightest
neutral Higgs scalar, of the pseudoscalar, and the minimal sum of the masses
of the lightest scalar and pseudoscalar Higgs bosons, respectively. The
minimum of the sum is significantly larger than the sum of the minima,
indicating that both Higgs masses cannot be minimized simultaneously.

\end{enumerate}

\end{document}